\documentclass{ws-procs10x7}
\usepackage{balance}
\usepackage{epsfig}

\newcolumntype{d}[1]{D{.}{.}{#1}}

\def\rmm   {\ensuremath{\Delta R(\mu\mu)}}

\def\vdef #1{\expandafter\def\csname #1\endcsname}
\def\vuse #1{\csname #1\endcsname}

\vdef{cut:default:ptlo}    {\ensuremath{  4.0 } } 
\vdef{cut:default:pthi}    {\ensuremath{100.0 } } 
\vdef{cut:default:etalo}    {\ensuremath{ -2.4 } } 
\vdef{cut:default:etahi}    {\ensuremath{  2.4 } } 
\vdef{cut:default:rmmlo}    {\ensuremath{  0.3 } } 
\vdef{cut:default:rmmhi}    {\ensuremath{  1.2 } } 
\vdef{cut:default:tip}      {\ensuremath{0.100 } } 
\vdef{cut:default:ptbs}    {\ensuremath{  5.0 } } 
\vdef{cut:default:chi2}    {\ensuremath{  1.0 } } 
\vdef{cut:default:l3d}    {\ensuremath{0.015 } } 
\vdef{cut:default:lxy/sxy}    {\ensuremath{18.0 } } 
\vdef{cut:default:lxy}    {\ensuremath{-0.100 } } 
\vdef{cut:default:cosalpha}    {\ensuremath{0.995 } } 
\vdef{cut:default:isoveto}    {\ensuremath{1000.0 } } 
\vdef{cut:default:isolation}    {\ensuremath{0.85 } } 
\vdef{cut:default:isocone}    {\ensuremath{  1.0 } }

\vdef{mBgSigma:s0}{\ensuremath {36.0 } }
\vdef{mBg1s:s0}  {\ensuremath {32.1 } }
\vdef{mBg1n:s0} {\ensuremath {0.08 } }
\vdef{mBg2s:s0}  {\ensuremath {60.2 } }
\vdef{mBg2n:s0} {\ensuremath {0.03 } }

\vdef{eAllCutsFact:s0}  {\ensuremath{{0.019 } } }
\vdef{eAllCutsFactE:s0} {\ensuremath{{0.002 } } }
\vdef{nAllCutsFact:s0}  {\ensuremath{{  6.1} } }
\vdef{nAllCutsFactE:s0}  {\ensuremath{{  0.3} } }

\vdef{nMassAllCutsFact:m0}  {\ensuremath{{ 13.8} } }
\vdef{nMassAllCutsFactE:m0} {\ensuremath{{ 22.0} } }
\vdef{eAllCutsFact:m0}  {\ensuremath{{2.74\times 10^{-7} } } }
\vdef{eAllCutsFactE:m0}  {\ensuremath{{2.96\times 10^{-8} } } }

\vdef{Lumi:d0}    {\ensuremath{ {10.0 } } }

\vdef{bgError}    {\ensuremath{{160 } } }
\vdef{sgError}    {\ensuremath{{25 } } }

\vdef{nMassAllCutsFactE:s0} {\ensuremath{{ 0.6} } }
\vdef{nMassAllCutsFactE:m0} {\ensuremath{{ 22.0} } }

\vdef{ExpectedUpperLimit}  {\ensuremath{{1.4\times 10^{-8}} } }

\input symbols.tex
\makeindex
\begin{document}

\title{STUDY OF \bsmm\ in CMS}

\author{Urs~Langenegger$^*$ }

\address{Institute for Particle Physics, ETH Zurich, 8093 Zurich, Switzerland\\$^*$E-mail: ursl@phys.ethz.ch}


\twocolumn[\maketitle\abstract{We  present  a  Monte Carlo  simulation
  study  of measuring  the rare  leptonic  decay \bsmm\  with the  CMS
  experiment  at the  LHC.   The study  is  based on  a full  detector
  simulation  for  signal  and  background  events.   We  discuss  the
  high-level  trigger algorithm  and the  offline event  selection.  }
 ]

\section{Introduction}

In the standard  model (SM), the decay \bsmm\  has a highly suppressed
rate\cite{Buras:2003td} of $\cbf  = (3.42\pm0.54)\times 10^{-9}$ since
it  involves a $b  \to s$  transition and  requires an  internal quark
annihilation  which  further  suppresses  the decay  relative  to  the
electroweak `penguin' $b \to s \gamma$ decay.  In addition, the decays
are helicity  suppressed by factors of $m_{\ell}^{2}$.   To date these
decays  have  not  been  observed;  upper limits  on  these  branching
fractions    are    a   topic    of    frequent    updates   at    the
$B$-factories\cite{Chang:2003yy,Aubert:2004gm} (for the decay $\bdmm$)
and  the Tevatron.\cite{Abazov:2004dj,CDF:prelim}  Currently  the best
upper  limit  is  from  the  CDF  collaboration\cite{CDF:prelim}  with
$\cbf(\bsmm) < 1.0\times 10^{-7}$ at 95\% confidence limit.

\begin{figure}[!htb]
 \centerline{\psfig{file=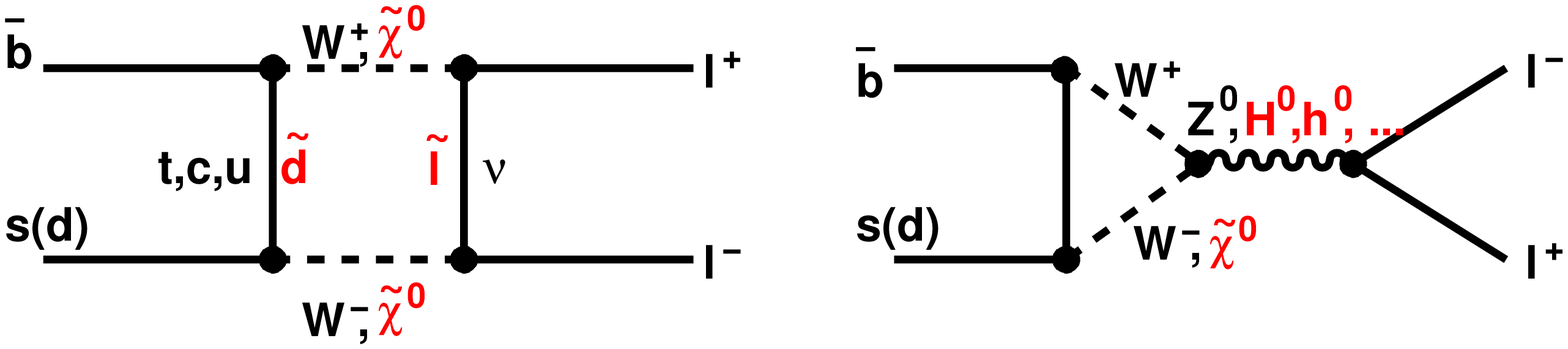,width=2.6in}}
 \caption{Feynman graphs for the decay \bsdmm\ illustrating possible
   new physics contributions.}
 \label{f:b2llnp}
\end{figure}

Since  these processes  are  highly  suppressed in  the  SM, they  are
potentially   sensitive  probes   of  physics   beyond  the   SM  (see
Fig.~\ref{f:b2llnp}).  In the  minimal supersymmetric extension of the
SM (MSSM) the branching fraction for these decays can be substantially
enhanced, especially at large $\tan \beta$.\cite{Babu:1999hn} For MSSM
with  modified  minimal flavor  violation  at  large $\tan\beta$,  the
branching  fraction  can  be  increased   by  up  to  four  orders  of
magnitude.\cite{Bobeth:2002ch} \bsdmm\ decays  can also be enhanced in
specific  models   containing  leptoquarks~\cite{Davidson:1993qk}  and
supersymmetric (SUSY) models without R-parity.\cite{Roy:1991qr}

There                  has                  been                  some
interest\cite{Kane:2003wg,Baek:2004tm,Dedes:2004yc} in using the decay
mode \bsmm\  to `measure' the  key parameter \tb\  of the MSSM  and to
constrain other  extensions of  the SM.  Lower  bounds on \tb\  can be
obtained    from   $\cbf(\bsmm)$   with    general   model-independent
assumptions.  Since \tb\ is also constrained from above due to general
principles, a lower bound is tantamount to a measurement of \tb.

\section{The CMS Experiment}

The Compact Muon Solenoid (CMS)  detector is well suited for the study
of leptonic $B$ decays. The  main components for this analysis are the
tracker and muon systems.

The CMS tracker  is an all-silicon detector, comprises  an inner pixel
vertex detector plus an outer strip track detector, and is immersed in
a   magnetic  field   of   4\Tesla.   The   pixel  detector   provides
high-precision measurements  of space points close  to the interaction
region  for  vertexing  and   effective  pattern  recognition  in  the
high-track multiplicity environment at  the LHC. The pixel detector is
composed of 1440 modules arranged  in three barrel layers (at a radial
distance of  $r = 4.4,\, 7.3,\,  10.2\cm$ from the  beampipe) and four
endcap  disks (at  $z =  \pm34.5,\, \pm46.5\cm$  from  the interaction
point).    The  barrel   detector   comprises  672   modules  and   96
half-modules,  the forward  detector is  built of  96 blades  with 672
modules.   With a  pixel  size of  $d_\phi\times  d_z =  100\mum\times
150\mum$,  a hit  resolution of  10--20\mum\ is  achieved.   The strip
tracker allows the precise  determination of charged particle momenta. 
It is  arranged in  the central  part as inner  (TIB) and  outer (TOB)
barrel, and  in the forward regions  as inner discs  (TID) and endcaps
(TEC).  The barrel part consists of 10  layers (4 in the TIB, 6 in the
TOB) and 12 layers  in the forward part (3 in the TID,  9 in the TEC). 
The  pitch  in the  strip  tracker  varies  between 80--180\mum.   The
material  inside  the active  volume  of  the  tracker increases  from
$\approx 0.4X_0$  at pseudo-rapidity  $\eta = 0$  to around  $1X_0$ at
$|\eta| \approx 1.6$, before decreasing to $\approx 0.6X_0$ at $|\eta|
= 2.5$.

The  CMS muon  system, incorporated  into the  magnet return  yoke, is
divided  into a  barrel ($|\eta|  < 1.2$)  and forward  parts  ($1.2 <
|\eta|  < 2.4$).   In the  barrel  region, where  the neutron  induced
background and  the muon rate is  small, drift tube  (DT) chambers are
used. In  the two endcaps cathode  strip chambers (CSC)  are deployed. 
In  addition, resistive  plate chambers  (RPC)  are used  both in  the
barrel and the  endcap region.  The RPC spatial  resolution is coarser
than for  the DT and CSC,  but their excellent  time resolution allows
the unambiguous identification of the correct bunch crossing.

\section{Event Samples}

Monte Carlo (MC) event samples were generated with {\sc Pythia~6.227}
and passed through a full detector  simulation based on {\sc Geant~4}. 
On  average  five pile-up  events  were  included,  appropriate for  a
luminosity of $\clu = 2\times10^{33} \cm^{-2}\sec^{-1}$.

Both signal  and background  MC event samples  have been  generated as
minimum bias  QCD events. In the  signal sample, \Bs\  mesons decay as
\bsmm.  The muons are required  to have transverse momentum $\pt^\mu >
3\gev$ and $|\eta^\mu| < 2.4$; the \Bs\ must have $\pt^{\Bs} > 5\gev$.

The  background sample  contains  two  muons with  $\pt  > 3\gev$  and
$|\eta|  <  2.4$.   Their  separation in  azimuth  and  pseudorapidity
$\rmm\equiv \sqrt{\Delta\phi^2 + \Delta\eta^2}$ is required to be $0.3
< \rmm <  1.8$.  Currently the background simulation  does not include
muons  due  to hadronic  in-flight  decays  or  punch-through.  It  is
estimated that  this hadronic  component will increase  the background
level    by    about    10\%.     Background    events    from    rare
$B_d,\,B_u,\,B_s,\,B_c,\,\Lambda_b$ decays are not included.

In total  20000 signal events  and about 15000 background  events have
been  analyzed.   The small  size  of  the  background sample  is  the
limiting factor of the present study.

\section{Trigger Strategy}

The  CMS detector  has a  twofold trigger  strategy.  The  first level
trigger,  with a latency  of $3.2\,\mu\sec$,  is based  on information
from  the  calorimeters  and  the  muon  system.   The  threshold  for
inclusive  isolated single  muons is  at $\pt^\mu>14\gev$,  for dimuon
events both  muons must have $\pt>3\gev$.  The  expected trigger rates
amount to  2.7\kHz\ and 0.9\kHz, respectively.  The  L1 trigger output
rate is at most 100\kHz.

The high-level trigger (HLT) is a software trigger, running on a large
processor  farm. The  HLT reduces  the overall  trigger rate  by three
orders  of magnitude  to about  100\Hz.  To  fit into  the  tight time
constraints imposed  by the  high input rate,  tracking at the  HLT is
sped up by  two concepts: (1) `Regional seed  generation' limits track
seeding to  specific regions  of interest, \eg,  a cone around  the L1
muon  candidate  direction.   (2)  `Partial  tracking'  pursues  track
reconstruction only until some criteria  are met, \eg, a resolution of
2\% on  the transverse momentum. Already with  six reconstructed hits,
both  the efficiency  and the  resolution are  comparable to  the full
tracking performance.

The HLT  strategy for  \bsmm\ proceeds along  the following  path. (1)
Verification of the two  L1 muon candidates.  (2) Tracks reconstructed
only with  the pixel detector are  used to compute a  list of possible
primary  vertices,  the  three  most significant  are  retained.   (3)
Regional  track reconstruction  with up  to six  hits is  performed in
cones around  the L1 muon  candidates.  (4) Reconstructed  tracks with
$\pt >  4\gev$ are paired and  retained if their  invariant mass falls
into predefined  regions for  the signal and  sidebands.  (5)  The two
tracks must have  opposite charge and are fit to  a common vertex; the
event  is  retained   only  when  the  fit  $\chi^2   <  20$  and  the
three-dimensional  flight  length   $l_{3d}  >  150\mum$.   With  this
selection,  the event  rate was  estimated\cite{Sphicas:2002gg}  to be
$<1.7\Hz$.

\section{Offline Analysis}

The offline analysis  selection focuses on a secondary  vertex that is
well  measured and separated  from the  primary vertex  and consistent
with the decay of an isolated \Bs\ meson.

The primary vertex is determined from all tracks with $\pt > 0.9\gev$,
because of the rather low-multiplicity track environment.

The   two    muons   must    have   opposite   charge,    $\pt^\mu   >
\vuse{cut:default:ptlo} $,  and $|\eta^\mu| < \vuse{cut:default:etahi}
$.   The azimuthal  and  pseudorapidity separation  of  the two  muons
$\vuse{cut:default:rmmlo} < \rmm < \vuse{cut:default:rmmhi} $ provides
a  powerful  reduction  of  gluon-gluon fusion  background  with  both
$b$-hadrons decaying semileptonically:  The muons of those $b$-hadrons
tend to be back-to-back, while  the signal shows a peaked distribution
with a maximum at $\rmm \sim 1$.

\begin{figure}[!htb]
 \centerline{\psfig{file=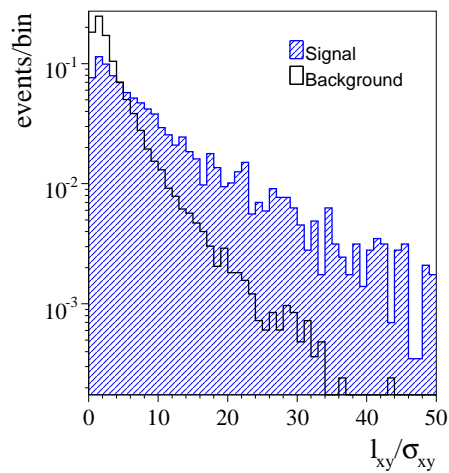,width=2.6in}}
 \caption{Decay length significance in
   the  transverse plane  for  signal and  background  MC events.  Both
   histograms are normalized to unity.}
 \label{f:selection}
\end{figure}

\Bs\ candidates are formed by  vertexing the two muon candidates.  The
vertex quality is required to  be $\chi^2 < \vuse{cut:default:chi2} $. 
The transverse momentum vector of  the \Bs\ candidate must be close to
the displacement of the secondary  vertex from the primary vertex: the
cosine  of the  opening angle  $\alpha$ between  the two  vectors must
fulfill   $\cos(\alpha)    >   \vuse{cut:default:cosalpha}   $.    The
significance  of the  \Bs\  candidate flight  length  $l_{xy}$ in  the
transverse  plane is defined  as $l_{xy}/\sigma_{xy}$  (illustrated in
Fig.~\ref{f:selection}),  where  $\sigma_{xy}$  is  the error  on  the
flight  length  with mean  $\langle\sigma_{xy}\rangle  = 120\mum$;  we
require $l_{xy}/\sigma_{xy} > \vuse{cut:default:lxy/sxy} $.

In  high-\pt\  gluon-splitting events  the  \bbbar\  quark pair  moves
closely together due to their boost, and the two decay vertices of the
resulting $b$-hadrons cannot be well separated in all cases.  However,
because of color reconnection, the hadronic activity around the dimuon
direction is  enhanced compared  to the signal  decay (where  only one
colorless  \Bs\  meson  decays).    This  is  exploited  in  isolation
requirements:  The  isolation  $I$   is  determined  from  the  dimuon
transverse momentum  and charged tracks  with $\pt>0.9\gev$ in  a cone
with half-radius  $r = \vuse{cut:default:isocone} $  around the dimuon
direction as $I \equiv \pt^{\mu\mu}/(\pt^{\mu\mu} + \sum_{trk}|\pt|)$;
we require $I > \vuse{cut:default:isolation} $.

\begin{figure}[!htb]
 \centerline{\psfig{file=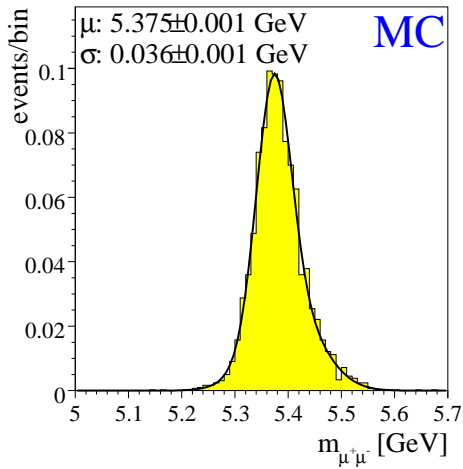,width=2.6in}}
 \caption{Dimuon mass distribution in signal MC events. The curve
   is a  fit of two  Gaussians, the displayed parameters  indicate the
   average mean and sigma. The histogram is normalized to unity.}
 \label{f:breco}
\end{figure}

Figure~\ref{f:breco} illustrates  the mass resolution  obtained on the
signal MC  event sample. The  distribution is fit with  two Gaussians,
the  quoted width  $\sigma  = \vuse{mBgSigma:s0}  \mev$ is  determined
according to
\begin{eqnarray*}
 \sigma^2 = \frac{N_n^2\sigma_n^2 + N_w^2\sigma_w^2}{N_n^2 + N_w^2},
\end{eqnarray*}
where $\sigma_n  = \vuse{mBg1s:s0} \mev $ ($\sigma_w =  \vuse{mBg2s:s0} \mev$)
and $N_n =  \vuse{mBg1n:s0} $ ($N_w = \vuse{mBg2n:s0} $)  are the width
and normalization of the narrow (wide) Gaussian, respectively.

Given  the  limited statistics  of  the  background  sample, no  event
remains after the application  of all selection requirements. However,
the absence of correlation  to the other selection requirements allows
a factorization  of the isolation  and $\chi^2$ requirements  from the
other  cuts in  the determination  of the  total  background rejection
factor.    The  dominant   sources  of   uncertainty  on   the  signal
($\pm\vuse{sgError}  \%$) and background  ($^{+\vuse{bgError} }_{-100}
\%$) yield are the statistical component of the background sample, the
impact   of  the   misalignment  on   the  transverse   flight  length
significance, and the assumption of factorizing cuts.

The total  selection efficiency for signal events  is $\varepsilon_S =
\vuse{eAllCutsFact:s0}   \pm   \vuse{eAllCutsFactE:s0}   $   and   the
background reduction factor is $\varepsilon_B = \vuse{eAllCutsFact:m0}
$.   With  this  selection,   the  first  $\vuse{Lumi:d0}  \invfb$  of
integrated  luminosity   will  yield  $n_S   =  \vuse{nAllCutsFact:s0}
\pm\vuse{nMassAllCutsFactE:s0}   $   signal    events   and   $n_B   =
\vuse{nMassAllCutsFact:m0}               ^{+\vuse{nMassAllCutsFactE:m0}
}_{-\vuse{nMassAllCutsFact:m0} } $ background  events in a mass window
of $m_{\Bs}  \pm 0.1\gev$.  With  this background estimate,  the upper
limit    on    the   branching    fraction    is   $\cbf(\bsmm)    \le
\vuse{ExpectedUpperLimit} $ at the 90\% C.L.

\section{Conclusions}

This study is limited by the size of the background MC sample.  In the
future,  it will  include  larger background  samples  and a  detailed
simulation of  rare $b$-hadron decays. The search  for \bsmm\ promises
an interesting start-up analysis  with the possibility of setting tight
constraints on  the MSSM.  With sufficient  integrated luminosity, the
precision  measurement  of  the  \bsmm\ branching  fraction  will  set
constraints on models of new physics.



\begin{thebibliography}{99}

  \bibitem{Buras:2003td}
 A.~J.~Buras,
 Phys.\ Lett.\ B {\bf 566}, 115 (2003).



  \bibitem{Chang:2003yy} M.~C.~Chang {\it et al.}  [BELLE Collaboration],
 Phys.\ Rev.\ D {\bf 68}, 111101 (2003).

  \bibitem{Aubert:2004gm}
 B.~Aubert {\it et al.}  [BABAR Collaboration],
 Phys.\ Rev.\ Lett.\  {\bf 94}, 221803 (2005).



  \bibitem{Abazov:2004dj}
 V.~M.~Abazov {\it et al.}  [D0 Collaboration],
 Phys.\ Rev.\ Lett.\  {\bf 94}, 071802 (2005).

 
  \bibitem{CDF:prelim}
 A.~Abulencia {\it et al.} [CDF collaboration], 
 CDF Public Note 8176 (2006).



  \bibitem{Babu:1999hn}
 K.~S.~Babu and C.~F.~Kolda,
 Phys.\ Rev.\ Lett.\  {\bf 84}, 228 (2000);
 S. R. Choudhury and N. Gaur, Phys. Lett. B 451, 86 (1999);
 C. S. Huang, \etal, Phys. Rev. D 63, 114021 (2001).


  \bibitem{Bobeth:2002ch}
 C.~Bobeth, T.~Ewerth, F.~Kruger and J.~Urban,
 Phys.\ Rev.\ D {\bf 66}, 074021 (2002).


  \bibitem{Davidson:1993qk}
 S.~Davidson, D.~C.~Bailey and B.~A.~Campbell,
 Z.\ Phys.\ C {\bf 61}, 613 (1994).

  \bibitem{Roy:1991qr}
 D.~P.~Roy,
 Phys.\ Lett.\ B {\bf 283}, 270 (1992).


  \bibitem{Kane:2003wg}
 G.~L.~Kane, C.~Kolda and J.~E.~Lennon,
 arXiv:hep-ph/0310042.

  \bibitem{Baek:2004tm}
 S.~Baek,
 Phys.\ Lett.\ B {\bf 595}, 461 (2004).

  \bibitem{Dedes:2004yc}
 A.~Dedes and B.~T.~Huffman,
 Phys.\ Lett.\ B {\bf 600}, 261 (2004).


\bibitem{Sphicas:2002gg}
  P.~Sphicas \etal\  [CMS Collaboration],
CERN-LHCC-2002-026.


\end{thebibliography}
\end{document}